\documentclass[a4paper,11pt]{article}

\usepackage[dvips]{graphicx}
\usepackage{amsfonts, color, float,  amsmath}
\usepackage{amssymb}
\usepackage{subfigure}
\usepackage{url}
\usepackage{array}

\usepackage{hyperref}

\usepackage{slashed}
\usepackage[svgnames]{xcolor} 

\usepackage[small,bf,sf]{caption} 
\usepackage{microtype}
\usepackage{cite}
\usepackage{authblk}
\usepackage[short]{datetime}
\usepackage[bottom]{footmisc} 
\usepackage{booktabs}  

\newdateformat{mydate}{\THEDAY \  \monthname \  \THEYEAR}

\allowdisplaybreaks

\newcommand{\Eq}[1]{Eq.~(\ref{#1})}
\newcommand{\Eqs}[2]{Eqs.~(\ref{#1}) and~(\ref{#2})}

\renewcommand{\Re}{{\rm Re}}
\renewcommand{\Im}{{\rm Im}}

\def\beq {\begin{equation}}
\def\eeq {\end{equation}}
\def\bea {\begin{eqnarray}}
\def\eea {\end{eqnarray}}

\newcommand{\ie}{{\it i.e.~}}

\newcommand{\gev}{\, \text{GeV}}

\AtBeginDocument{}

\title{\LARGE {\bf \sffamily \boldmath  Towards   tests of quark-hadron duality with functional analysis  and spectral function data}\vspace{0.3cm}}

\author[a]{Diogo Boito}
\author[b]{Irinel Caprini\vspace{0.3cm}}

\affil[a]{\it  Instituto de F\'isica de S\~ao Carlos, Universidade de S\~ao Paulo, CP 369, 13560-970, S\~ao
Carlos, SP, Brazil\vspace{0.3cm}}

\affil[b]{\it Horia Hulubei National Institute for Physics and Nuclear Engineering,\newline P.O.B. MG-6, 077125 Magurele, Romania}

\date{}

\begin{document}
\begin{flushright}
{
 \mydate \today
} 
\end{flushright}

\vspace*{-0.7cm}
\begingroup
\let\newpage\relax
\maketitle
\endgroup

\begin{abstract}
\noindent
 The presence of terms that violate quark-hadron duality in the expansion of QCD Green's  functions is a generally accepted fact.
 Recently, a new approach was proposed for the study of duality violations (DVs), which exploits the existence of a rigorous lower bound on the functional distance, measured in a certain norm, between a ``true" correlator and its approximant calculated theoretically along a contour in the complex energy plane.  In the present paper we pursue the investigation of functional-analysis based tests towards their application to real spectral function data. We derive a closed analytic expression for the minimal functional distance based on the general weighted $L^2$ norm and discuss its relation with the distance measured in $L^\infty$ norm. Using fake data sets obtained from a realistic toy model in which we allow for covariances inspired from the publicly available ALEPH spectral functions, we obtain by Monte Carlo simulations the statistical distribution of the strength parameter that measures the magnitude of the DV term added to the usual operator product expansion (OPE).  The results show that, if the region with large errors near the end-point of the spectrum in $\tau$ decays is excluded, the functional-analysis based tests using either $L^2$ or $L^\infty$ norms are able to detect, in a statistically significant way, the presence of DVs in realistic spectral function pseudodata.

\end{abstract}

\thispagestyle{empty}




\section{Introduction}

The presence of additional terms in the QCD Green's functions, beyond those generated by OPE (understood as perturbation theory and power corrections), is a generally accepted fact, with support both from theory and phenomenology. According to the standard terminology \cite{Shif, Blok, Shif1}, these terms are said to violate quark-hadron duality. We recall that, in its conventional sense, quark-hadron duality assumes that the description in terms of the OPE, valid on the Euclidian axis and at complex energies, can be analytically continued to match with the description in terms of hadrons, which live on the Minkowski axis.

DVs are supposed to arise from contributions of internal lines with soft momenta in the Feynman diagrams, which are not included in the OPE. Their calculation from first principles is, at least at present, impossible.  Quantitative understanding must be based on realistic models, whose main features have been tested against experimental data.  Two types of specific mechanisms have been suggested, one considering quarks in an instanton field \cite{Shif, Blok, Shif1}, the other based on narrow-resonance saturation in the large-$N_c$ limit \cite{Shif1, GPPR02, Cata1, Cata2, Cata3}. Both mechanisms are materialized in exponentially suppressed terms on the spacelike axis, which exhibit oscillations when analytically continued to the timelike axis.  More formal arguments in favour of the existence of DVs are provided by ideas of resurgence and the associated trans-series \cite{MStrans}. The assumption that the OPE,  expected to be a divergent expansion with increasing large-order coefficients, is actually an asymptotic series, leads also naturally to the presence of additional exponentially suppressed contributions \cite{Peris}.  However, beyond these somewhat general arguments, no detailed dynamical calculation of the additional contributions present in the theoretical expression of the Green's functions is available.

The phenomenological extraction of DVs is far from trivial because one must detect terms exponentially suppressed as 
function of energy, while an infinity of terms logarithmically and power suppressed, {\em i.e.} larger in principle, are 
neglected in the standard truncated expansions of the Green functions. Since these expansions are actually divergent, the 
magnitude of the neglected terms can be quite substantial.  Moreover, as mentioned above, the confrontation between 
theory and experiments implies an analytic continuation in the complex energy plane, with its known instabilities and 
pitfalls. Analyticity is usually exploited by means of a Cauchy integral relation along a contour in the complex plane for the 
QCD polarization amplitude of interest multiplied by a suitable weight. This allows one to build sum-rules that relate the 
integrated theoretical predictions on the contour to weighted integrals over the spectral function data on the positive 
Minkowski axis.  The weight is chosen such as to enhance or to suppress the contribution of various terms in the theoretical 
expression of the amplitude.  The impact of DVs for practical calculations is therefore sensitive to the weights that are 
employed and vary depending on the quantity of interest.  When extracting QCD parameters, for example, different weight 
functions have been advocated. In some works, DVs are explicitly taken into account by means of realistic 
parametrizations~\cite{GAPP10,GAPRS16,Boito0,Boito1,Boito2,Boito3}, which allows for a quantitative control of their 
contribution, while in others DVs are ignored on the basis of their suppression by the weight functions employed~
\cite{ALEPH,DHSS15,PRS16}.  The reliability of the different approaches is still being questioned~\cite{PRS16,BGMP16} 
and, therefore, a better understanding of DVs would certainly contribute to the precision with which QCD parameters can 
be extracted. This is particularly true for the determination of the coupling $\alpha_s$ from the $\tau$-hadronic spectral 
functions.

 In the recent paper \cite{CGP14}, a method based on functional analysis was proposed in order to test the presence of DVs in QCD. The method starts from the obvious remark that the ``true" polarization amplitude  and its approximate theoretical expression are entirely different functions, with different analytic properties,  which cannot coincide  in the complex energy plane. Moreover,  defining a functional distance, measured in a certain norm, between  these two functions along a contour in the complex plane,  a rigorous nonzero lower bound on this distance can be shown to exist. In particular, for the functional distances defined in $L^\infty$ and in $L^2$ norms, the lower bound can be calculated by an explicit algorithm involving the QCD approximant in the complex plane and  an infinity of Fourier coefficients  obtained from the spectral function  (``moments") measured experimentally on a part of the timelike axis.

 As argued in Ref.~\cite{CGP14}, the minimal distance between the true function and its approximant can be used as a tool for detecting the presence of DVs.  In particular, from the variation of the minimal distance with respect to a parameter $\mu$ that measures the strength of the duality violating contribution, one can infer the optimal value of this parameter.  Formulated in this way, the problem becomes analogous to the search for new physics beyond the standard model (SM) in experiments at very high energies, where one tests for the presence of new physics through a ``strength parameter'' $\mu$ of the signal, while treating SM as a background.  In our case, the ``new physics'' is represented by DV terms, while OPE is the background representing the ``known physics''.

The application of these ideas to a toy model proposed in Ref.~\cite{Cata2} indicated that the new approach might be useful for detecting the presence of DVs in QCD.  The asymptotic expansion of the exact model contains, besides a purely perturbative term and an expansion identified with higher-dimension terms in the OPE, an additional term that can be interpreted as a DV contribution. The minimal functional distance defined in Ref.~\cite{CGP14}, calculated with the spectral function of the model and a truncated OPE to mimic the physical cases, displayed a sharp minimum at the true value of the strength $\mu$ of the DV term. In particular, the functional distance measured in $L^\infty$ norm proved to be more sensitive to the variation of the parameter $\mu$ than the distance measured in $L^2$ norm. However, the effect of the experimental uncertainties inherent in the spectral function used as input was only barely touched in Ref.~\cite{CGP14}. A detailed investigation of this aspect is crucial for assessing the usefulness of the method to detect DVs from real data.  In the present paper we address precisely this problem.

We consider the same toy model proposed in Ref.~\cite{Cata2}, assuming now that the spectral function is measured only in a finite number of bins with uncertainties and correlations similar to those reported in real experiments on hadronic $\tau$ decays. It turns out that a statistical interpretation of the minimal distances defined by functional analysis is difficult to assess theoretically. Therefore, we perform an empirical study based on pseudodata, where fake data on the spectral function are generated in a number of bins, with a multivariate Gaussian distributions with covariances inferred from the ALEPH covariance matrix for the vector ($V$) channel~\cite{ALEPH}. The statistical distribution of the parameter $\mu$ that measures the magnitude of the DV term added to the usual OPE is then derived by Monte Carlo simulations, allowing the extraction of a mean and a standard deviation. The aim is to establish if the method is able to detect, in a statistically significant way, the presence of DVs from error-affected experimental measurements. We also compare the procedures based on $L^\infty$ and $L^2$ norms and establish which is the most eficient tool when the uncertainties in the spectral function are taken into account. In the process, we give closed analytical expressions for the functional distances in a generalized weighted $L^2$ norm that interpolates almost exactly between $L^2$ and $L^\infty$.

The plan of the paper is as follows. We start, in Sec.~2, with a brief review of the approach proposed in Ref.~\cite{CGP14}, defining the minimal functional distances in $L^\infty$ and $L^2$ norms and presenting the algorithms for their calculation.  Sec.~\ref{sec:math} contains two new mathematical developments important for applications: in subsection ~\ref{sec:approx} we prove that the minimal distance based on the general weighted $L^2$ norm can be written down in a closed analytic form, and in subsection~\ref{sec:approx} we derive a suitable approximation of the minimal distance based on $L^\infty$ norm by a class of weighted $L^2$ norms.  In Sec.~\ref{sec:toy} we briefly review the toy model and describe the data generation with ALEPH-based covariances. Sec.~\ref{sec:results} contains our main results and Sec.~\ref{sec:conc} is devoted to our conclusions.


\section{Theoretical framework}
\label{sec:review}

We begin with a short presentation of the work performed in  Ref.~\cite{CGP14}.
The main idea is to quantify the difference, along a contour in the complex $s$ plane,  between the 
QCD description $\Pi_{\rm QCD}(s)$ of a correlator of light quark currents and its true value  $\Pi(s)$. By QCD description one understands the perturbative part, the contribution from OPE condensates and possible duality violations:
\beq
\Pi_{\rm QCD}(s) = \Pi_{\rm OPE}(s) + \Pi_{\rm DV}(s),
\label{eq:PiQCD}
\eeq
where $\Pi_{\rm OPE}$ encompasses both the purely perturbative contribution (or dimension $D=0$ contribution) and the power corrections.

\begin{figure}[!t]
\begin{center}
\includegraphics[width=.5\columnwidth,angle=0]{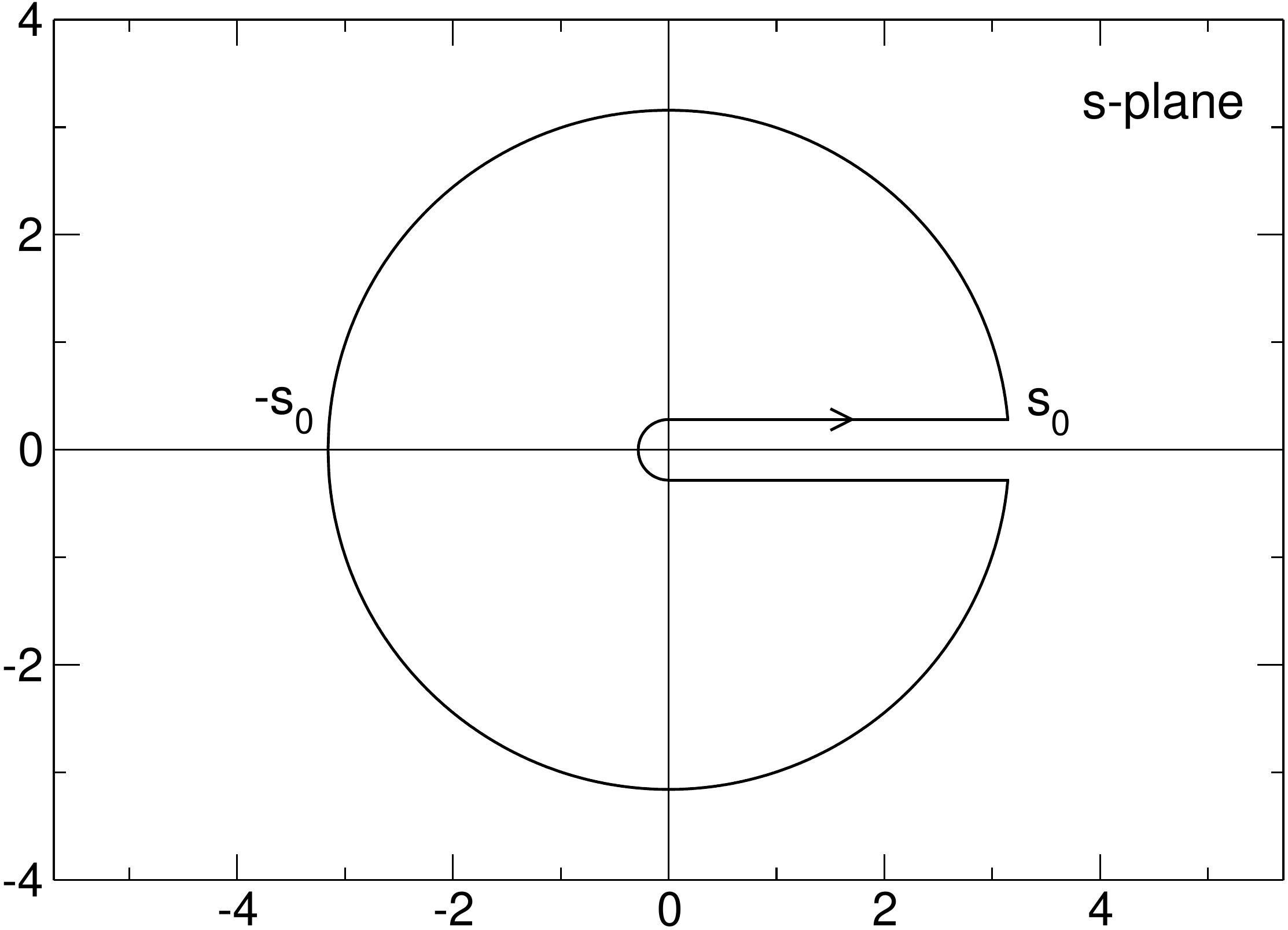}
\caption{The contour in the complex s-plane.}
\label{fig:circle}
\end{center}
\end{figure}

For definiteness the contour was taken as the circle $|s|=s_0$ shown in Fig. \ref{fig:circle}, where $s_0$ is  sufficiently large such that the QCD approximant is valid.
Measuring the distance by the $L^\infty$ norm \cite{Duren}, we consider the quantity
\beq\label{eq:delta}
\delta = {\sup_{\theta \in (0,2\pi)}} |\Pi(s_0 e^{i \theta}) - \Pi_{\rm QCD}(s_0 e^{i\theta})|,
\eeq
where  $\Pi(s)$ is the ``true", physical function, known to be an analytic function  in the $s$-plane cut along the real axis for $s\ge 4m_\pi^2$, which satisfies the Schwarz reflection condition $\Pi(s^*)=\Pi^*(s)$. In addition, its discontinuity 
\begin{equation}
\mbox{Im}\,\Pi(s+i\epsilon)=\sigma(s)
\end{equation}\label{eq:sigma}
 is known  experimentally on a limited energy range,  $4m_\pi^2\leq s\leq s_0$.  

The exact value $\delta_{\rm exact}$ of $\delta$ cannot be computed in QCD for lack of the true  $\Pi(s)$: the properties stated above do not specify uniquely the function, but define a whole class of admissible functions to which the physical one must belong.  If we define
\beq\label{eq:delta0}
  \delta_0 = \min_{\{\Pi\}} \sup_{\theta \in (0,2\pi)} |\Pi(s_0 e^{i \theta}) - \Pi_{\rm QCD}(s_0 e^{i\theta})|,
\eeq
where the minimization is with respect to all functions $\Pi$ in this admissible class, then it follows that  
\beq\label{eq:delta0lim}
\delta_{\rm exact} \ge  \delta_0.
\eeq

As shown in Ref.~\cite{CGP14}, the quantity $\delta_0$ can be calculated by applying a duality theorem in functional optimization \cite{Duren}. For completeness, we present below the main steps of the proof. We  make first the simple change of variable
\begin{equation}\label{eq:z}
z\equiv \tilde z(s)=\frac{s}{s_0},
\end{equation}
which maps the domain shown in Fig.~\ref{fig:circle} onto the unit disk $|z|\leq 1$ cut along a segment  of the real axis.  Various classes of analytic functions have been defined in the canonical domain $|z|\leq 1$. In particular, adopting the functional distance (\ref{eq:delta0}) we are lead   naturally to the class $H^\infty$ of functions $F(z)$ analytic inside the disk and bounded on the circle $|z|=1$, with the
 $L^\infty$ norm defined as the supremum of the modulus
along the boundary $\vert z\vert=1$: 
\begin{equation}\label{eq:Hinf}
\Vert F\Vert_{L^\infty}\equiv \sup_{\theta\in (0,2\pi)}\vert F(e^{i\theta}) \vert.
\end{equation} 
We consider also the class $H^2$ of  functions analytic inside the disk and of finite $L^2$ norm on the  frontier $|z|=1$, where
\begin{equation}\label{eq:H2}
\|F\|_{L^2}\equiv \left[\frac{1}{ 2\pi}\int_0^{2\pi}\vert F(e^{i\theta})\vert^2 d\theta\right]^{1/2},
\end{equation}
and the more general  class $H^2_w$ of analytic functions of finite $L^2_w$ norm, where $L^2_w$ is the more general norm defined as 
\begin{equation}\label{eq:H2w}
\|F\|_{L^2_w}\equiv \left[\frac{1}{ 2\pi}\int_0^{2\pi}\vert w(e^{i\theta})F(e^{i\theta})\vert^2 d\theta\right]^{1/2},
\end{equation}
in terms of a weight $w(\zeta)$ given on the boundary $\zeta= e^{i\theta}$ of the unit circle.

As shown in Ref.~\cite{CGP14}, the problem (\ref{eq:delta0}) can be written in the equivalent form
\begin{equation}\label{eq:inf1}
\delta_0=\min_{g\in H^\infty} \Vert g-h\Vert_{L^\infty}.
\end{equation}
Here the minimization is performed with respect to all the functions $g(z)$
analytic  in the disk $\vert z \vert<1$ and bounded on the frontier, and  $h$ is a known complex
 function defined on the boundary  $\zeta=\exp(i \theta)$ of the unit circle by
\beq\label{eq:h}
h(\zeta) = -\frac{\zeta}{\pi} \int_{x_\pi}^{1+\eta}dx \frac{\sigma(s_0 x)}{x(x-\zeta -i \epsilon)} + \Pi_{\rm QCD}(s_0\zeta),
\eeq
where $\eta>0$ is an arbitrary parameter introduced for technical reasons, which does not appear in the final result (see \cite{CGP14} for more explanations).

 The solution of the problem (\ref{eq:inf1}) has been  obtained in Ref.~\cite{CGP14} by means of a duality theorem in functional optimization \cite{Duren}. This theorem reads
\begin{equation}\label{eq:dual1}
\min_{g\in H^\infty}\|g-h\|_{L^\infty}=\sup_{w,f \in S^2}\left\vert \frac{1}{ 2\pi}
\oint\limits_{\vert \zeta\vert =1} w(\zeta)f(\zeta)h(\zeta) d\zeta\right\vert,
\end{equation}
where the functions  $w(z)$ and $f(z)$  belong to the unit sphere $S^2$ of $H^2$, {\em i.e.}  are analytic in $|z|<1$ and satisfy the conditions
\begin{equation}\label{eq:wfnorm}
\Vert w\Vert_{L^2} \leq 1\,,\quad \quad\quad\Vert f\Vert_{L^2}\leq 1\,,
\end{equation}
where the norm is defined in Eq.~(\ref{eq:H2}).

We recall that all the functions considered here are real analytic,  {\em i.e.} satisfy the reflection property written as $f(z^*)=f^*(z)$.
 Therefore, if one writes the Taylor expansions
 \begin{equation}\label{eq:wf}
w(z)=\sum_{n=0}^\infty w_nz^n\,,\quad \quad f(z)= \sum_{m=0}^\infty f_m z^m\,,
\end{equation}
 coefficients will be real and, due to (\ref{eq:wfnorm}), will satisfy the conditions
\begin{equation}\label{eq:wfl2}
\sum_{n=0}^\infty w_n^2\leq 1\,,\quad\quad \quad\sum_{m=0}^\infty f_m^2\leq 1\,.
\end{equation} 

The supremum in the right-hand side of Eq.~(\ref{eq:dual1}) can be calculated by means of a rather  simple  numerical algorithm, as shown in Ref.~\cite{CGP14}.  Namely, let ${\cal H}$ be the Hankel matrix defined as
\begin{equation}\label{eq:hank}
{\cal H}_{nm}=c_{n+m-1}\,, \quad n,m\geq 1\,,
\end{equation}
in terms of the real coefficients
\beq
c_n = \frac{1}{\pi} \int_0^1 dx\, x^{n-1}\sigma(s_0x)  + \frac{1}{2\pi}\int_0^{2\pi}d\theta\, e^{in\theta} \Pi_{\rm QCD}(s_0 e^{i\theta}),\quad n \geq 1\,.\label{eq:cn}
\eeq
The coefficients $c_n$ defined in Eq.~(\ref{eq:cn}) are actually the negative-frequency Fourier coefficients which measure the ``non-analytic'' part  in $|z|<1$ of the complex function $h(\zeta)$ defined in Eq.~(\ref{eq:h}).  One may recognize in them the moments used in traditional finite-energy sum rules based on a Cauchy integral relation for $\Pi$ multiplied with a power of $s$ along the contour of Fig. \ref{fig:circle}. Then, $\delta_0$ is obtained as the spectral norm
\begin{equation}\label{eq:del0H}
\delta_0=\Vert{\cal H}\Vert_{L^2}=\Vert{\cal H}\Vert\,,
\end{equation}
\ie~the square root of the greatest
 eigenvalue of the positive-semidefinite matrix ${\cal H}^\dagger {\cal H}$.

 In the numerical  calculations, the matrix  ${\cal H}^\dagger {\cal H}$ is truncated  at a finite order 
$m=n=N$, using the fact that for large $N$ the successive approximants tend toward the exact result (for a formal proof of convergence see 
Appendix E of Ref.~\cite {CiNe} and for numerical tests see Ref.~\cite{CaSa}).
 By the duality theorem, the initial
functional minimization problem~(\ref{eq:delta0}) is thus reduced to a rather simple numerical computation.

One can define also the minimal functional distance $\delta_2$ based on the $L^2$ norm 
\beq\label{eq:delta2}
  \delta_2 = \min_{\{\Pi\}} \left[\frac{1}{2 \pi} \int_0^{2 \pi} d\theta |\Pi(s_0 e^{i \theta}) - \Pi_{\rm QCD}(s_0 e^{i\theta})|^2\right]^{1/2},
\eeq
which can be written in the $z$ variable as
\begin{equation}\label{eq:infL2}
\delta_2=\min_{g\in H^2} \Vert g-h\Vert_{L^2},
\end{equation}
for the same function $h$ defined in Eq.~(\ref{eq:h}). 
The solution of this minimization problem has the simple form \cite{CGP14}
\beq
\delta_2 = \left[ \sum_{n=1}^\infty c_n^2  \right]^{1/2},\label{eq:L2Norm}
\eeq
in terms of the same coefficients $c_n$ defined in Eq.~(\ref{eq:cn}).

More generally, we consider the functional distance based on the more general norm $L^2_w$ defined in Eq.~(\ref{eq:H2w}), when instead of (\ref{eq:infL2}) we must solve the problem
\begin{equation}\label{eq:infL2w}
\delta_{2,w}=\min_{g\in H^2} \Vert w(g-h)\Vert_{L^2}\,,
\end{equation} where $w$ is a suitable weight. It can be shown,  without loss of generality, that $w$ can be taken as the boundary value of an {\em outer} function  \cite{Duren}, \ie  a function $w(z)$ analytic and without zeros in $|z|<1$. It is easy to show then that the solution of the problem (\ref{eq:infL2w}) is
\beq
\delta_{2,w} = \left[ \sum_{n=1}^\infty d_n^2  \right]^{1/2}, \label{eq:L2Normw}
\eeq
where the real numbers $d_n$ are the weighted moments
\beq\label{eq:dn}
d_n=\frac{1}{\pi}\int\limits_0^{1} x^{n-1} w(x)\sigma(s_0 x) dx +\frac{1}{2\pi}\int\limits_0^{2 \pi} e^{i n\theta} w(e^{i \theta} ) \Pi_{\rm QCD}(s_0 e^{i \theta} )d\theta\,,
\eeq
depending on the  function $w(z)$. The quantity  $\delta_2$ defined in the standard
 $L^2$ norm is obtained from these relations for $w(z)\equiv 1$.

In practice, as in the calculation of $\delta_0$ by means of Eq.~(\ref{eq:del0H}), the infinite sums in Eqs.~(\ref{eq:L2Norm}) and (\ref{eq:L2Normw})
are truncated after a finite number of terms and the convergence  towards the values $\delta_2$ and $\delta_{2,w}$ is tested numerically. 
Actually, as we will show in the next section, the infinite summation in the general case (\ref{eq:L2Normw}) can be performed exactly and the
minimal distance $\delta_{2,w}$ can be written in a closed analytic form.


\section{New mathematical developments}
\label{sec:math}

\vspace{0.3cm}

\subsection{\boldmath Closed analytic form of $\delta_{2,w}$}\label{sec:analyt}
A compact analytic form for the quantity $\delta_{2,w}$ can be obtained easily by performing the summation upon $n$ in the expression (\ref{eq:L2Normw}). For convenience we write the real coefficients $d_n$ defined in Eq.~(\ref{eq:dn}) as
\beq\label{eq:dnsum}
d_n=d_{n,\sigma}+d_{n, {\rm QCD}},
\eeq
where the significance of the terms is obvious. Then we obtain from  (\ref{eq:L2Normw}):
\beq\label{eq:sum}
\delta_{2,w}^2=\sum_{n=1}^\infty d_{n,\sigma}^2 +2\sum_{n=1}^\infty d_{n,\sigma} d_{n, {\rm QCD}} + \sum_{n=1}^\infty d_{n, {\rm QCD}}^2.
\eeq

Using the expression of  $d_{n,\sigma}$ from  (\ref{eq:dn}), the first sum in Eq.~(\ref{eq:sum}) can be written immediately as
\beq\label{eq:s1}
\sum_{n=1}^\infty d_{n,\sigma}^2=\frac{1}{\pi^2}\int_0^1 dx \int_0^1 dy \frac{w(x) w( y) \sigma(s_0 x) \sigma(s_0 y)}{1-x y}. 
\eeq
The second sum in Eq.~(\ref{eq:sum}) is written in a convenient form by using for the coefficients $d_{n, {\rm QCD}}$ the expression
\beq
d_{n, {\rm QCD}}=\frac{1}{4\pi}\int_0^{2 \pi}\left[ e^{i n\theta}w( e^{i \theta} ) \Pi_{\rm QCD}(s_0 e^{i \theta} ) + e^{-i n\theta} w^*( e^{i \theta} ) \Pi_{\rm QCD}^*(s_0 e^{i \theta} )\right]d\theta\,,
\eeq
which is explicitly real.  Using further the reality property of the functions $w$ and $\Pi_{\rm QCD}$, {\em i.e.} the relation  $w(z^*)\Pi_{\rm QCD}(s^*)=w^*(z) \Pi^*_{\rm QCD}(s)$, the integration interval can be reduced to $(0,\, \pi)$. Thus we obtain after a straightforward calculation 
\bea\label{eq:s2}
&& 2\sum_{n=1}^\infty d_{n,\sigma} d_{n, {\rm QCD}} =
\frac{2}{\pi^2}\int_0^1 dx w(x) \sigma(s_0 x)\\&&\times\int_0^{\pi} d\theta
\frac{\Re \left[w(e^{i \theta})\Pi_{\rm QCD}(s_0 e^{i \theta})\right](\cos\theta-x) -\Im \left[w(e^{i \theta})\Pi_{\rm QCD}(s_0 e^{i \theta})\right]\sin\theta}{1-2 x\cos\theta+x^2}.\nonumber
\eea
 We note that the end singularities  at $x=y=1$  in the integrand of (\ref{eq:s1}) and at $x=1$, $\theta=0$ in the integrand of (\ref{eq:s2}) are logarithmically integrable.

The last sum in Eq.~(\ref{eq:sum}) can be written as
\bea\label{eq:s3}
 \sum_{n=1}^\infty d_{n, {\rm QCD}}^2 \hspace{-0.5cm}&&= \frac{1}{4 \pi}\int_0^{2\pi} d\theta |w( e^{i \theta}) \Pi_{\rm QCD}(s_0 e^{i \theta})|^2\\
\hspace{-0.5cm}&&-\frac{P}{8 \pi^2}\int_0^{2\pi} d\theta \int_0^{2\pi} d\theta'  |w( e^{i \theta})\Pi_{\rm QCD}(s_0 e^{i \theta}) w( e^{i \theta'})\Pi_{\rm QCD}(s_0 e^{i \theta'})|\nonumber\\
\hspace{-0.5cm}&&\times \,\,\frac{\sin[\Phi(\theta)-\Phi(\theta')+\frac{\theta-\theta'}{2}]}{\sin\frac{\theta-\theta'}{2}}\nonumber,
\eea
where $P$ denotes the principal part and
\beq
\Phi(\theta)=\arg[w( e^{i \theta})\Pi_{\rm QCD}(s_0 e^{i \theta})]
\eeq
is the phase of the complex function $w(z) \Pi_{\rm QCD}(s)$ on the circle $|s|=s_0$, \ie on $|z|=1$. 

For the numerical evaluation it is more convenient to write the second term in the r.h.s. of Eq.~(\ref{eq:s3}) in the equivalent form:
\bea\label{eq:s3alt}
&&-\frac{P}{8 \pi^2}\int_0^{2\pi} d\theta \int_0^{2\pi} d\theta'   \left(\Re\left[w( e^{i \theta})\Pi_{\rm QCD}(s_0 e^{i \theta}) w^*( e^{i \theta'})\Pi^*_{\rm QCD}(s_0 e^{i \theta'})\right] \right.\nonumber\\
&&+ \left.\Im \left[w( e^{i \theta})\Pi_{\rm QCD}(s_0 e^{i \theta}) w^*( e^{i \theta'})\Pi^*_{\rm QCD}(s_0 e^{i \theta'})\right] \cot\frac{\theta-\theta'}{2} \right).
\eea
By collecting the terms in Eqs.~(\ref{eq:s1}), (\ref{eq:s2}), (\ref{eq:s3}) and (\ref{eq:s3alt}) we obtain the final expression of the squared of $\delta_{2,w}$
\bea\label{eq:delta2w}
&&\delta_{2,w}^2=\frac{1}{\pi^2}\int_0^1 dx \int_0^1 dy \frac{w(x) w( y) \sigma(s_0 x) \sigma(s_0 y)}{1-x y}
+ \frac{2}{\pi^2}\int_0^1 dx w(x) \sigma(s_0 x)\nonumber\\
&&\times\int_0^{\pi} d\theta
\frac{\Re \left[w(e^{i \theta})\Pi_{\rm QCD}(s_0 e^{i \theta})\right](\cos\theta-x) -\Im \left[w(e^{i \theta})\Pi_{\rm QCD}(s_0 e^{i \theta})\right]\sin\theta}{1-2 x\cos\theta+x^2}\nonumber\\
&&+ \frac{1}{4 \pi}\int_0^{2\pi} d\theta |w( e^{i \theta}) \Pi_{\rm QCD}(s_0 e^{i \theta})|^2\\
&&-\frac{P}{8 \pi^2}\int_0^{2\pi} d\theta \int_0^{2\pi} d\theta'   \left(\Re\left[w( e^{i \theta})\Pi_{\rm QCD}(s_0 e^{i \theta}) w^*( e^{i \theta'})\Pi^*_{\rm QCD}(s_0 e^{i \theta'})\right] \right.\nonumber\\
&&+ \left.\Im \left[w( e^{i \theta})\Pi_{\rm QCD}(s_0 e^{i \theta}) w^*( e^{i \theta'})\Pi^*_{\rm QCD}(s_0 e^{i \theta'})\right] \cot\frac{\theta-\theta'}{2} \right)\nonumber.
\eea

 All the integration intervals can be further reduced to $(0, \,\pi)$ by taking into account, as explained above, the reality property of the functions,  which implies that the real (imaginary) parts are even (odd) functions of $\theta$.

\subsection{ \boldmath Approximation of $L^\infty$ norm by a suitable class of  $L^2_w$ norms}\label{sec:approx}
 We show now that it is possible to approximate the minimal distance $\delta_0$ measured by the $L^\infty$ norm by a class of minimal distances $\delta_{2,w}$ defined by the weighted $L^2_w$ norms.  We follow an argument put forward for the first time in  Refs. \cite{CD1980, Cap1981}, which is based on the duality theorem Eq.~(\ref{eq:dual1}) applied for solving the original minimization problem  (\ref{eq:inf1}).

 We note that  the r.h.s. of Eq. (\ref{eq:dual1}) requires the calculation of the supremum upon two sets of functions, $w(z)$ and $f(z)$,  which are analytic in the unit disk $|z|<1$ and of $L^2$ norm  bounded by 1. 
 The idea is to calculate first the
supremum upon one class of functions, say $f$, keeping the
other one fixed.  We note that the r.h.s. of Eq. (\ref{eq:dual1}) can be written as
\beq\label{eq:}
\left\vert \frac{1}{ 2\pi}
\oint\limits_{\vert \zeta\vert =1} w(\zeta)f(\zeta)h(\zeta) d\zeta\right\vert =\left\vert\sum\limits_{n=1}^\infty d_n f_{n-1}\right\vert,
\eeq
where  $f_n$ are the Taylor coefficients defined in Eq.~(\ref{eq:wf}) and  $d_n$ are negative-frequency Fourier coefficients of the product $w(\zeta) h(\zeta)$, given by the weighted moments  (\ref{eq:dn}). Then   Eq. (\ref{eq:dual1}) becomes
\beq
\delta_0=\sup_{\{w\}} \sup_{\{f_n\}}\left\vert \sum\limits_{n=1}^\infty d_n f_{n-1}\right\vert.
\eeq
 The supremum upon the coefficients $f_n$ subject to the second condition (\ref{eq:wfl2}) can be evaluated immediately by Cauchy-Schwarz inequality, leading to
\beq\label{eq:LinfL2}
\delta_0=\sup_{\{w\}} \left[\sum\limits_{n=1}^\infty d_n^2\right]^{1/2},
\eeq
where the dependence of the coefficients $d_n$ on the weight $w$ is given in Eq.~(\ref{eq:dn}). Finally, by using 
(\ref{eq:L2Normw}), we write this relation as
\beq\label{eq:d0d2}
\delta_0=\sup_{\{w\}} \delta_{2,w}.
\eeq
We emphasize that this is an exact relation, which states that the minimal distance $\delta_0$ in the $L^\infty$ norm is the largest value from the class of distances $\delta_{2,w}$ in the weighted $L^2_w$ norm, for all the weight $w$ subject to the first condition (\ref{eq:wfnorm}).

Of course, the problem is not yet solved, we still have to calculate the supremum in (\ref{eq:d0d2}). The procedure makes sense if one can find a suitable,
simple parametrization of the functions $w$, such that the maximization  upon this limited class approximates well  the exact $\delta_0$. It turns out that such a
choice exists \cite{CD1980, Cap1981}: the main observation is
that one can obtain approximately the maximum modulus of a function on a certain interval by computing the normalized integral of its modulus squared in a variable that dilates the region where the modulus of the function reaches its maximum. Therefore, one can approximate the $L^\infty$ norm (\ref{eq:Hinf}) of an arbitrary function
 by an $L^2$ norm (\ref{eq:H2}) defined on the integration range 
distorted by a suitable change of variable. In order to obtain it, we consider the conformal mapping of
the unit disc $|z|\leq 1$ onto itself, achieved by the so-called Blaschke transformation \cite{Duren}
\beq\label{eq:blas}
z' = \frac{ z - a}{  1 - a^* z}, 
\eeq
where $a$ is an arbitrary parameter with $|a|<1$. Since we consider real analytic functions, one can restrict $a$ to real values. The transformation (\ref{eq:blas}) maps in particular the unit circle $|z|= 1$ onto itself.
This change of variable in the $L^2$ norm (\ref{eq:H2}) introduces the Jacobian $|dz'/dz|$, which corresponds to a weight function $w(z)$ in the weighted $L^2_w$ norm (\ref{eq:H2w}), of the form
\beq\label{eq:w}
w(z)=\frac{\sqrt{1-a^2}}{1- a z},
\eeq
where $a$ is a real parameter in the range $(-1, 1)$. It is easy to check that this function satisfies the first condition (\ref{eq:wfnorm}).

By the above remark, the functional supremum in Eq.~(\ref{eq:d0d2}) was reduced to a maximization with respect to a single real parameter $a$. The minimal distance $\delta_0$ can thus be calculated approximately by a relatively simple algorithm: first one calculates  the minimal distance $\delta_{2,w}$ given in  (\ref{eq:L2Normw}), with the particular choice (\ref{eq:w}) of the weight. Then the parameter $a$ is varied in the range $(-1,1)$ and the largest value of  $\delta_{2,w}$ is retained. This problem is numerically quite simple, especially since, as shown in the previous subsection,  $\delta_{2,w}$ for an arbitrary weight $w$ can be written in an analytic compact form.

Some hints on the optimal value of the parameter $a$ are obtained from the  specific properties of the input. Thus, we note that for values of $a$ close to 1, the function $w(z)$ is large near $z=1$, \ie  near $s=s_0$, both on the circle and on the real axis. Therefore, in this case the weighted  norm  (\ref{eq:infL2w}) is dominated by the region of the circle near the timelike axis. One can expect that such values of $a$ would be useful in order to detect DVs that are large only near the timelike axis. 
We shall test these expectations in the numerical studies reported in Sec. \ref{sec:results}.

\section{Toy data generation}
\label{sec:toy}

The main goal of the formal developments presented in the previous
sections is to provide tools for the validation of DV models using
information from the spectral function data. It is therefore
instrumental to test the procedure with toy data sets generated from a
realistic model for which the DVs are known exactly. Part of the work
described here is an extension of analytical results of
Ref.~\cite{CGP14} to a more realistic situation, where the spectral
function comes in the form of a binned data set subjected to
statistical fluctuations.  With the application to ALEPH data in mind,
we shall consider data sets that are obtained from a realistic
covariance matrix. In this section we discuss the central model used for the
exercises performed in this work as well as how we construct our
covariance matrix.  With these two ingredients, we have full control over
the problem and can perform simulations in order to understand how the
procedure can be applied to real data.

We start with a brief review of the model that we employ for this
exercise.\footnote{An extended discussion of the model in the present
  context can be found in Ref.~\cite{CGP14}.} The model was
introduced in Ref.~\cite{Cata2}, based on previous ideas from
Refs.~\cite{Shif1,Blok,GPPR02}. To be concrete, here we focus on the vector spectral
function. The description  is based on a ``Regge tower'' of resonances and upon
including the $\rho$ meson pole into the tower, the correlator $\Pi(s)$
assumes the following exact form
\beq
\Pi_{\rm model}(s)=-\frac{1}{\zeta}  \frac{2 F^2}{\Lambda^2}\psi\left(\frac{v+m_0^2}{\Lambda^2}\right),\label{eq:PiModel}
\eeq
where we defined
\beq
v=\Lambda^2\left(\frac{-s-i\epsilon}{\Lambda^2}\right)^{\zeta},
\eeq
and $\psi(v)=\Gamma'(v)/\Gamma(v)$ is the Euler digamma function. We employ the following  set
of parameters:
\beq
\zeta = 0.95,\qquad F=133.8\,\rm{MeV}, \qquad \Lambda = 1.189\,{\rm GeV}, \qquad m_0 = 0.75\,{\rm GeV},
\eeq
which provides, for our purposes, a realistic description of the experimentally observed
vector spectral function of the QCD correlator.

The asymptotic expansion of the digamma function can be used in order
to obtain an OPE-type description of the correlator $\Pi_{\rm model}(s)$. 
Truncating the expansion at an order $N_{\rm OPE}$ it reads
\beq
\Pi_{\rm OPE}(s) =-\ \frac{2 F^2}{\Lambda^2}\, C_0 
\log\left(\frac{-s}{\Lambda^2}\right)\ +\sum_{k=1}^{N_{\rm OPE}}\frac{C_{2k}}{v^k}.
\label{eq:PiOPE}
\eeq
The first term corresponds to the ``purely perturbative'' part and the other
terms are power corrections, akin to the condensate contributions  of QCD. The explicit expression of the coefficients
that appear in  $\Pi_{\rm OPE}$ are
\beq
C_0=1, \qquad  C_{2k}=\frac{2}{\zeta} (-1)^k \frac{1}{k}
\Lambda^{2k-2} F^2 B_k\left(\frac{m_0^2}{\Lambda^2}\right), \quad k\ge 1,
\eeq
with $B_k(x)$ representing Bernoulli polynomials.

The asymptotic expansion of Eq.~(\ref{eq:PiOPE}) is not accurate
near the timelike axis, as in the case for the OPE in QCD. For large
enough $|s|$ and $\Re( s)>0$ the description can be improved taking into
account the DVs. In practice, the DV term can be obtained from the
reflection property of the digamma function~\cite{Cata2,CGP14}. The
following modified approximant is thus obtained
\beq \label{eq:OPEDV}
 \Pi_{\rm model}(s)\approx 
\Pi_{\rm OPE}(s)+\Pi_{\rm DV}(s)\,,
\eeq
valid for large enough $|s|$ and for $\Re(s)>0$. The DV contribution
is given in the first quadrant ($\Im( s)>0$ and $\Re(s)>0$) by
\beq\label{eq:PiDV}
\Pi_{\rm DV}(s)=\frac{2 \pi F^2}{\Lambda^2\zeta}\left[-i+
\cot\left[\pi\left(\frac{-s}{\Lambda^2}\right)^{\zeta}+\pi\,\frac{m_0^2}{\Lambda^2}\right]\right],
\eeq
and can be defined in the lower half-plane using  Schwarz reflection as  $\Pi_{\rm DV}(s^*)=\Pi^*_{\rm DV}(s)$. For $\Re(s)\leq 0$ this correction is assumed to vanish.

Comparing the modulus of the exact function, \Eq{eq:PiModel}, along
the upper semi-circle $s=s_0 e^{i\theta}$, $\theta\in (0, \pi)$, with
its approximants, \Eqs{eq:PiOPE}{eq:OPEDV}, one learns that the
truncated OPE-type expansion of Eq.~(\ref{eq:PiOPE}) provides an
accurate description except close to the timelike axis ($\theta=0$),
as expected in QCD.  The addition of the DV term fixes this deficiency
and the approximated description of \Eq{eq:OPEDV} becomes excellent also in the
vicinity of the timelike axis. (We refer to Ref.~\cite{CGP14} for a
visual account of this comparison.)

For the numerical exercises described in this work we use the model of
Eq.~(\ref{eq:PiModel}) as our central description. Hence, the OPE for the model
and the DV contribution are exactly known and are given by \Eqs{eq:PiOPE}{eq:PiDV}, respectively. The  values
of the vector spectral function for toy data generation are obtained,
thus, from
\beq
\sigma_{\rm model}(s) =  \Im\, \Pi_{\rm model}(s+i\varepsilon),\qquad s\in [0, m_\tau^2].
\label{eq:SFModel}
\eeq
In order to mimic the experimental situation, the interval
$[0:m_\tau^2]$ is split in  $N_{\rm b}$ bins and the
central value of each bin is obtained from a statistical distribution
that fluctuates  around the
values of Eq.~(\ref{eq:SFModel}) calculated at the center of each bin. We turn now to the issue of the
covariance matrix that governs these fluctuations.

Our toy data generation is performed having in mind the application to
the ALEPH spectral functions~\cite{ALEPH}. It is therefore desirable that the
covariances used reflect those of ALEPH data sets. One could simply
adopt the ALEPH covariances as such, since they are publicly available
\cite{ALEPHData}, and generate toy data sets following a statistical
distribution given by this matrix, together with the central values of
Eq.~(\ref{eq:SFModel}). The price to pay is that one would have to use
the ALEPH binning of the interval $[0:m_\tau^2]$. In the most recent
version of the data sets, due to an improved unfolding procedure, an
adaptive binning was used which results in bins with different widths,
notably with larger bins towards the edge of the spectrum.  Here we
prefer to adopt a fixed bin width, for simplicity, and we choose $N_{\rm b}$ such as to
have more bins than ALEPH towards the end-point of the spectrum.  
This allows us to have a
finer description at higher energies. The
accompanying realistic covariances are obtained from a numerical
interpolation of the ALEPH covariance matrix for the vector
channel. In this way, we preserve a fixed binning together with a
covariance matrix that has all the main properties of ALEPH's, namely,
strong correlations between neighbouring bins, larger uncertainties
towards the end-point of the spectrum and, of course, uncertainties
that are of the same order of those of ALEPH's data.

\begin{figure}[!ht]
\begin{center}
\includegraphics[width=.9\columnwidth,angle=0]{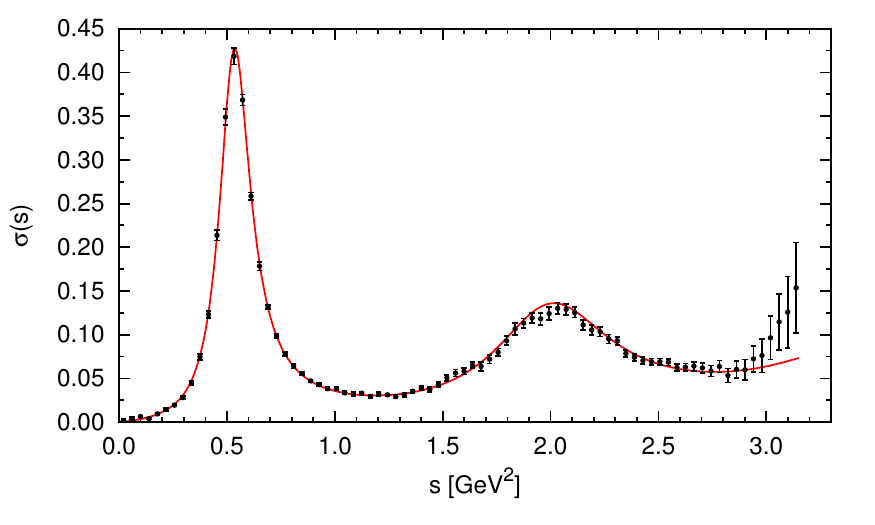}
\caption{An example of a toy data set obtained from the central values of the model given in Eq.~(\ref{eq:SFModel}) with covariances  from a numerical interpolation of ALEPH's covariance matrix for the vector channel~\cite{ALEPHData}. The solid line gives the central value of the model for comparison.}
\label{fig:DataExample}
\end{center}
\end{figure}

In the present work we adopt $N_{\rm b}
=80$ (which is in line with what is used in the experimental analyses
\cite{Opal,Aleph05,ALEPH}), the central values of Eq.~(\ref{eq:SFModel}), and
the covariance matrix obtained from a numerical interpolation of the
ALEPH covariances obtained from \cite{ALEPHData}, as described above. Toy data sets can then be
generated from a multivariate Gaussian distribution.  An example of a
data set generated in this way is displayed in Fig.~\ref{fig:DataExample}. We show also the central values of Eq.~(\ref{eq:SFModel}) for comparison. In
this figure, the strong correlations are clearly visible, mainly
towards the end-point of the spectrum, where the uncertainties are
also larger.

When using data sets for the calculation of the functional distances $\delta$
discussed in Secs.~\ref{sec:review} and \ref{sec:math}, weighted
integrals over the spectral function such as those entering Eq.~(\ref{eq:cn}) must be discretized. We are going
to adopt integration by rectangles, as is usual when dealing with this
type of integrals of the spectral
functions~\cite{ALEPH,Boito3}. However, weight functions $w_n$ with high
powers of the energy variable appearing, for example, in 
  Eq.~(\ref{eq:cn}),  vary strongly within a bin. It is therefore
necessary to average over the weight function inside a bin to improve
the numerical result.\footnote{In the case of ALEPH data this prescription is sometimes used  due to the large bin widths of the right-most bins~\cite{Boito3}.} 
The numerical counterpart of a typical integral reads then
\beq
\int_0^{s_0} ds\, w_n(s/s_0)\sigma(s) \approx \sum_{i=1}^{[s_0]}  \sigma (\bar s_i)\int_{\bar s_i-\Delta_b/2}^{\bar s_i+\Delta_b/2}ds \, w_n(s/s_0),\label{eq:BetterIntegral}
\eeq
where $\bar s_i$ is the value of $s$ at the center of the $i$th bin,
$\Delta_b$ is the fixed bin width, and $[s_0]$ represents the index of the
last included bin --- here we always work with $s_0$ values that
correspond to the right edge of a bin. The same was applied for the calculation of the
 relevant integrals which appear in the analytic form of $\delta_{2,w}$ derived in Sec. \ref{sec:analyt}.   We have tested that this
algorithm provides enough accuracy for the explorations performed in
this work.

\section{Results}
\label{sec:results}

 We apply now the functional-analysis based tools  to test in practice the description of DVs. To illustrate the potential of the method, a useful approach is  to introduce in the approximate description (\ref{eq:OPEDV}) of the correlator a strength parameter $\mu$ that allows one to tune the contribution of the DVs. Formally, we do this by using for $\Pi_{\rm QCD}$ in the formalism presented in Secs. \ref{sec:review} and \ref{sec:analyt}, instead of Eq. (\ref{eq:PiQCD}), the more general expression
\beq 
\label{eq:Pimu}
\Pi_{\rm QCD}(s) = \Pi_{\rm OPE}(s) + \mu \Pi_{\rm DV}(s),
\eeq
where the true value  of the strength parameter is $\mu=1$. As in  Ref.~\cite{CGP14}, to simulate the situation of the light-quark correlators in QCD,  we take $\Pi_{\rm OPE}$ as the asymptotic  expansion (\ref{eq:PiOPE}) of the exact model truncated after $N_{\rm OPE}=5$ terms. For $\Pi_{\rm DV}$ we take the prediction (\ref{eq:PiDV}) of the  model.  

In Ref.~\cite{CGP14}, it was shown by means of analytical computation  that $\delta_0$ has a sharp minimum at the correct value $\mu=1$,  when one employs the description of  DVs that follows from
the model used for $\Pi(s)$. The alternative quantity $\delta_2$ displayed also a minimum at
 the correct value of $\mu$, but this minimum was found to be shallower~\cite{CGP14}. In this section we investigate the impact of the use of   spectral function data with realistic covariances to the above findings. It will be interesting to make use of the weighted $L^2_w$ norm, since it permits a continuous and almost exact  interpolation between the $L^2$ and the $L^\infty$ norms, as well as a study of other weighted norms, such as ``pinched" norms. The analytical results obtained for $\delta_{2,w}$ shall also be instrumental in this analysis.


\subsection{\boldmath Comparison between $L^\infty$ and $L^2_w$ norms }\label{sec:comp}

 We check first on the toy model the approximation of the minimal distance $\delta_0$ based on  $L^\infty$ norm by the distances $\delta_{2,w}$ based on the norms $L^2_w$, using the particular class of weights given in (\ref{eq:w}).  In this discussion we use, as in Ref. \cite{CGP14}, the exact spectral function of the model, with no errors, and the OPE expansion truncated at $N_{\rm OPE}=5$.

\begin{figure}[!ht]
\begin{center}
\includegraphics[width=.75\columnwidth,angle=0]{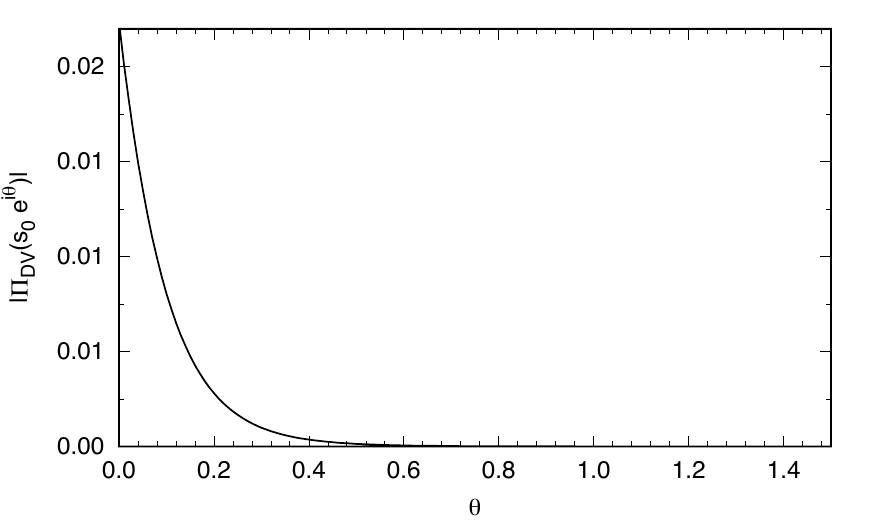}
\caption{Modulus of $\Pi_{\rm DV}(s)$ as a function of $\theta$ on the first quadrant of the circle $s=s_0 \exp (i \theta)$ with $s_0=2.76\,  \gev^2 $.  $\Pi_{\rm DV}(s)$ is zero in the left half of the $s$ complex plane. }
\label{fig:DV}
\end{center}
\end{figure}

From Fig. \ref{fig:DV},  which shows the modulus of $\Pi_{\rm DV}(s)$ as a function of $\theta$ on the first quadrant of the circle $s=s_0 \exp (i \theta)$, one can see that the DV part of the model is strongly peaked towards the Minkowskian axis. Therefore, following the discussion at the end of Sec. \ref{sec:approx},   a weight strongly peaked towards $s=s_0$ (\ie $\theta=0$) is expected to give the best approximation of $L^\infty$ norm by weighted $L^2_w$ norms for this model. As shown below, the expectation is confirmed.

\begin{figure}[H]
\begin{center}
\includegraphics[width=.75\columnwidth,angle=0]{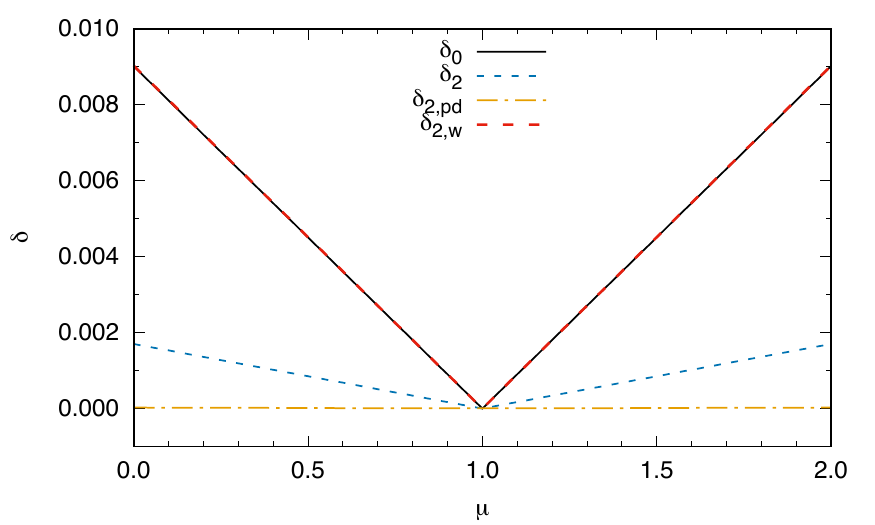}
\caption{Variation of the minimal functional distances with $\mu$, for $s_0=2.76\,  \gev^2 $. $\delta_0$ is based on $L^\infty$ norm, $\delta_{2,w}$, which coincides practically with $\delta_0$, is obtained with $w(z)$ of the form (\ref{eq:w}) for $a=0.96$, and $\delta_{2,{\rm pd}}$ is obtained with the pinched weight (\ref{eq:wpd}).  The calculation of the norms is done with the exact spectral function $\sigma(s)$ of the model, using $N=100$ Fourier coefficients.  }
\label{fig:LinfL2}
\end{center}
\end{figure}

In Fig. \ref{fig:LinfL2} we show the variation with $\mu$ of several functional distances, calculated with the algorithms based on Fourier coefficients truncated at $N=100$. We set in this exercise the radius of the circle in Fig. \ref{fig:circle} to $s_0=2.76\,\gev^2$,  but the results are similar for other choices, including $s_0=m_\tau^2$.

As usual,  $\delta_0$ denotes the minimal distance measured in $L^\infty$ norm, calculated from the norm (\ref{eq:del0H}) of the Hankel matrix, Eq.~(\ref{eq:hank}). For calculating $\delta_{2,w}$, we used the truncated sum (\ref{eq:L2Normw}), with the expression (\ref{eq:dn}) of $d_n$ and a weight $w$ of the form (\ref{eq:w}) with $a=0.96$. This weight drastically dilates the region near $\theta=0$ on the circle, increasing its contribution to the $L^2_w$ norm.
One can see that, for this choice of the weight, the  distance  $\delta_{2,w}$ practically coincides with $\delta_0$.   Both curves are steeper than the standard distance $\delta_{2}$, which corresponds  to the weight $w=1$,  as already remarked for $\delta_0$ in Ref.~\cite{CGP14}. The figure shows also that the minimal distance $\delta_{2,{\rm pd}}$, calculated with a pinched weight of the form\footnote{One refers as ``pinched" to weight functions that have a zero for $s=s_0$.} 
\beq\label{eq:wpd}
w_{\rm pd}=\left(1-\frac{s}{s_0}\right)^2,
\eeq
is much less sensitive to the variation of $\mu$, which is not surprising since this type of weight suppresses the region where the DV term is nonzero. 

The optimal value of the parameter $a$ was found empirically, by computing $\delta_{2,w}$ for several values of $a$ close to 1, and keeping the value leading to the best approximation of $\delta_0$. Of course, the best value achieving the supremum in (\ref{eq:LinfL2}) depends  also on the other ingredients of the input. Thus, for a different number $N$ of Fourier coefficients taken into account in the calculation of the norms a slightly different value of $a$ might yield the best approximation. Also, a slightly different  optimal value of $a$ is expected if the input spectral function is slightly changed. This remark will be useful for understanding the results of the simulations performed below, which take into account the uncertainties on the spectral function.

\begin{figure}[!ht]
\begin{center}
\includegraphics[width=.75\columnwidth,angle=0]{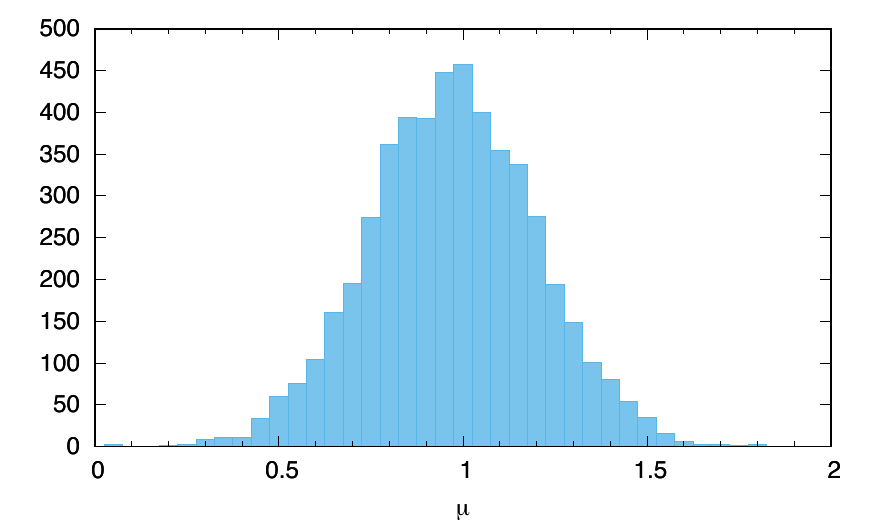}
\caption{Typical $\mu$ distribution obtained from Monte Carlo simulations.}
\label{fig:muHist}
\end{center}
\end{figure}

\subsection{Stability  and comparison with the summed results}\label{sec:stabil}

We start the simulations by investigating the computation of the distance $\delta_2$ measured with $L^2$ norm from the truncated version of Eq.~(\ref{eq:L2Norm}) --- which can be viewed as a special case of  Eq.~(\ref{eq:L2Normw})  with the appropriate choice of the weight $w$.  For a given data set, one can compute the value of $\delta_2$, or more generically, of $\delta_{2,w}$, after a truncation of the infinite sums at the $N$-th term. 
The value of the norm can then be minimized numerically with respect to the strength parameter $\mu$. Due to the statistical fluctuations, each toy data set that is generated yields a different value of $\mu$. We repeat this procedure for 5,000 different data sets, in a reproducible way, in order to obtain the statistical distribution of the parameter $\mu$. The final 
value of $\mu$ can be read off from the distributions. We quote central values given by the  medians and uncertainties defined by $68$\% confidence levels, but the distributions are, to a very  good approximation, Gaussian, as illustrated with the histogram shown in Fig. \ref{fig:muHist}.

In Table~\ref{tab:NDep}, we show the dependence of the best values of $\mu$ on the number $N$ of included terms in the truncated version of Eq.~(\ref{eq:L2Norm}). In this table we  choose $s_0=2.76$~GeV$^2$, which avoids some of the bins with larger uncertainties (see Fig.~\ref{fig:DataExample}). One can conclude from this table that the convergence of the results seems to be satisfactory; with a few hundred terms in the sum the results are already stable.  Furthermore, the exercise indicates that with a realistic data set the error in $\mu$ is such that we are able to detect the presence of DVs, \ie  $\mu=1$,  from  their absence, $\mu=0$,  in a statistically meaningful way.

\begin{table}[!ht]
\begin{center}{
\caption{Values of $\mu$ from the minimization of $\delta_2$ as a function of the number $N$ of coefficients included in the truncated sum (\ref{eq:L2Norm})  for $s_0 = 2.76\gev^2$ . The numerical integrals needed for the $c_n$ are computed as in Eq. (\ref{eq:BetterIntegral}).  The central  $\mu$ value is obtained from the median of the distribution. }\vspace{2mm}
\begin{tabular}{cc}
\toprule
$N$ & $\mu$\\
\midrule
5   & $1.1 \pm 1.4 $   \\
15  & $1.00\pm 0.21$   \\
100 & $1.00\pm 0.22$   \\
150 & $1.00 \pm 0.23$  \\
300 & $1.00 \pm 0.23$  \\
500 & $1.00 \pm 0.23$  \\
\bottomrule
\end{tabular}
\label{tab:NDep}
}\end{center}
\end{table}

We now turn to the dependence of the predictions on the choice of $s_0$. The fact that the last few bins suffer from  a much larger uncertainty, combined with the decrease of the  DV contribution at  higher  $s$, has the consequence that the choice of larger $s_0$ produces less precise determinations of $\mu$. In Table~\ref{tab:s0Dep} we compare values of $\mu$ obtained from $\delta_0$ and $\delta_2$ from the truncated versions of Eqs.~(\ref{eq:del0H}) and~(\ref{eq:L2Norm}), respectively,  for different values of $s_0$ (we use $N=300$ terms in the sums).  Two main conclusions can be drawn from this table. First, as expected, the uncertainties are larger when $s_0$ is chosen to be closer to the edge of the spectrum. For $\delta_2$, the determination loses statistical significancy rather fast when  the last few bins are included, and at $s_0=m_\tau^2$ one can no longer distinguish in a meaningful way the central value $\mu=1.0$ from the absence of DVs. Second, the use of $\delta_0$ leads to broader $\mu$-distributions and hence to
larger uncertainties. This effect is small for lower $s_0$ values and the results are essentially undistinguishable from those obtained using $\delta_2$. At $s_0=m_\tau^2$, however, the uncertainty is the double of the $\delta_2$ counterpart.\footnote{In order to obtain the results of Table 2 for $s_0=m_\tau^2$ it becomes important to allow for negative central values in the toy data spectral functions. These are rare, but do occur for the last few bins where the uncertainties, following the recent ALEPH reanalysis, are rather large. } We conclude that the deeper minimum of $\delta_0$ with respect to $\mu$ observed in the analytical calculations of Ref.~\cite{CGP14} (and seen also above in  Fig. \ref{fig:LinfL2}) does not translate into a narrower $\mu$-distribution when the errors on the spectral function are taken into account. 
 
\vspace{0.2cm}
\begin{table}[!t]
\begin{center}{
\caption{Values of $\mu$ from the minimization of $\delta_2$ and $\delta_0$ for different values of $s_0$. $N=300$ terms are included in the sum (\ref{eq:L2Norm})  and in the  Hankel matrix used in  (\ref{eq:del0H}). The numerical integrals needed for the $c_n$ are computed as in Eq. (\ref{eq:BetterIntegral}). The central  $\mu$ value is obtained from the median of the distribution. }\vspace{3mm}
\begin{tabular}{lll}
\toprule
 $s_0$ & $\mu$ from $\delta_2$ & $\mu$ from $\delta_0$\\
\midrule
$2.76$~GeV$^2$   & $1.00\pm 0.23$   &    $1.00 \pm 0.26$  \\
$2.84$~GeV$^2$   & $1.00\pm 0.29$   &    $1.01 \pm 0.30 $    \\
$3.00$~GeV$^2$    & $1.01 \pm 0.56$  &    $1.01 \pm 0.70 $    \\
$m_\tau^2$       & $1.0 \pm 1.6$   &    $1.0 \pm 3.0 $    \\
\bottomrule
\end{tabular}
\label{tab:s0Dep}
}\end{center}
\end{table}

A final validation of the results obtained in Tabs.~\ref{tab:NDep} and~\ref{tab:s0Dep} can be obtained using the closed analytical form of $\delta_{2,w}$ derived in Sec.~\ref{sec:analyt}.  The use of the weighted norm  $L^2_w$ is particularly  convenient as it allows for an almost exact interpolation between the $L^2$ and $L^\infty$ norms and, at the same time, is amenable to a fully analytical treatment of the minimization problem. 

\begin{table}[!ht]
\begin{center}{
\caption{Optimal values of $\mu$ from the minimization of the exact analytic expression of $\delta_{2,w}$, for two values of $s_0$. Three weights $w$ are used: $w=1$ corresponding to the standard $L^2$ norm, the weight (\ref{eq:w}) with $a=0.96$ and the pinched weight (\ref{eq:wpd}). The results are obtained with 5,000 toy data sets, the representative $\mu$ value being the median of the distributions.  }
\vspace{3mm}
\begin{tabular}{llll}
\toprule
$s_0$ & $w=1$\hspace{1cm} & $w=\frac{\sqrt{1-a^2}}{1- a s/s_0}$  \hspace{0.3 cm} &  $w=\left(1-\frac{s}{s_0}\right)^2$ \\
\midrule
2.76\gev$^2$ & $1.00\pm 0.23$  & $1.00\pm 0.26$   & $1.00\pm 0.70 $     \\
$m_\tau^2$ & $1.0\pm 1.6$  &  $1.0\pm 2.6$  &  $1.0 \pm 2.7$    \\
\bottomrule
\end{tabular}
\label{tab:analytic}
}\end{center}
\end{table}

Using the decomposition (\ref{eq:Pimu}) of $\Pi_{\rm QCD}$, we can write  Eq. (\ref{eq:delta2w}) as a quadratic polynomial of $\mu$ of the form 
\beq\label{eq:delta2wmu}
\delta_{2,w}^2 = b_0+ \left[b_1 +\frac{2}{\pi^2}\int_0^1 w(x) \sigma(s_0 x) F_{\rm DV}(x)dx \right]\mu +b_2 \mu^2,
\eeq
where 
\beq\label{eq:FDV}
 F_{\rm DV}(x)=\int\limits_0^{\pi} d\theta
\frac{\Re \left[w( e^{i \theta})\Pi_{\rm DV}(s_0 e^{i \theta})\right](\cos\theta-x) -\Im \left[w( e^{i \theta})\Pi_{\rm DV}(s_0 e^{i \theta})\right]\sin\theta}{1-2 x\cos\theta+x^2},
\eeq
and the calculable coefficients $b_i$ can be read off from  (\ref{eq:delta2w}). In particular, it is easy to see that only $b_0$  depends on the spectral function, the coefficients $b_1$ and $b_2$ involving only the values of the theoretical expressions $\Pi_{\rm OPE}(s)$ and $\Pi_{\rm DV}(s)$  on the circle $|s|=s_0$. 

The  optimal value of  $\mu$, which achieves the minimum of (\ref{eq:delta2wmu}), is obtained in a straightforward  manner as
\beq\label{eq:muopt}
\mu=-\frac{1}{2 b_2}\left[b_1+\frac{2}{\pi^2}\int_0^1 w( x) \sigma(s_0 x) F_{\rm DV}(x)dx\right].
\eeq 
 This formulation turns out to be very convenient for numerical simulations, which can be done  directly on the spectral function, avoiding the calculation of many experimental moments and the issue of truncating infinite sums. In practice, the coefficients $b_1$ and $b_2$ and the function $F_{\rm DV}$ are calculated only once, being fixed during data generation. This amounts to a considerable reduction of the computational time required by the statistical simulations. 

 In Tab.~\ref{tab:analytic} we display the ranges of $\mu$ obtained from simulations using the analytic result (\ref{eq:muopt}), for $s_0=2.76\gev^2$ and $s_0=m_\tau^2$ and three choices for the weight: $w=1$, which corresponds to the standard norm $L^2$ with the  minimal distance  $\delta_2$, the expression (\ref{eq:w}) with $a=0.96$, expected to approximate well the $L^\infty$ norm, and the pinched weight (\ref{eq:wpd}).
 We remark the perfect agreement between the results  quoted in Table \ref{tab:analytic} for the weight $w=1$ and the values obtained from  $\delta_2$  in Table~\ref{tab:s0Dep} for the same values of $s_0$.  This validates the convergence of the results  based on truncated sums of Fourier coefficients. We also remark a very good agreement between the results from $\delta_0$ and those from its approximated version in the second column of Tab.~\ref{tab:analytic}. (We discuss these results further in the next section.)

As seen from  Table ~\ref{tab:analytic},  in all cases the central value of the parameter $\mu$ coincides with the true central value. This result was expected, having in view the  precise theoretical input used along the circle in our study. The confirmation of this expectation is a test of the numerical algorithms used in the calculations. In particular, the integral in (\ref{eq:muopt}) had to be computed using the improved algorithm described in (\ref{eq:BetterIntegral}), \ie the product $w(x) F_{\rm DV}(x)$ was integrated exactly over each bin. 
On the other hand, the uncertainties quoted in the various entries of Table ~\ref{tab:analytic} are quite different.
 The explanation of these results  and their relevance for the application of the method to real data will be discussed in the next subsection.

\subsection{Discussion}\label{sec:weights}
The two weights, used in the simulations with the analytic form of  $\delta_{2,w}$ beside $w=1$, are quite extreme: the expression  (\ref{eq:w}) with  $a=0.96$ strongly enhances the contribution of the region near the point $s=s_0$ on the circle shown in Fig. \ref{fig:circle}.  Since, as seen from Fig.~\ref{fig:DV}, the magnitude of $\Pi_{\rm DV}$  is strongly peaked near the timelike axis,  the corresponding minimal distance $\delta_{\rm 2,w}$  will be very sensitive to the variation of the strength parameter $\mu$, as seen from Fig.~\ref{fig:LinfL2}. The same figure shows also that for this weight the norm $L^2_w$ approximates well the  $L^\infty$ norm. On the contrary, the weight (\ref{eq:wpd}) suppresses the region near $s=s_0$, which explains the low sensitivity of the corresponding distance $\delta_{\rm 2,pd}$ to the variation of  $\mu$, visible in  Fig.~\ref{fig:LinfL2}.

The above remarks refer to a fixed spectral function $\sigma(s)$. When this quantity is varied within errors during the simulations, the two extreme weights respond in a different way. The weight (\ref{eq:w}), which enhances also the region near $s=s_0$ on the real axis, will be more sensitive to the variation of the input data, since the errors are larger towards the end of the spectrum. For the lower value $s_0=2.76\gev^2$, when the errors are still moderate, the  effect of the variation of the input data turns out to be comparable to the opposite effect produced by the larger sensitivity to the variation of $\mu$. As a consequence, the spread of the $\mu$-distribution  for $w$ of the form  (\ref{eq:w}) with  $a=0.96$ is comparable to that obtained with $w=1$, as seen from the second and third columns of the first row of Table \ref{tab:analytic}.  On the other hand,  the pinched  weight $w_{\rm pd}$ ensures a low sensitivity of $\delta_{\rm 2,pd}$  to the variations of the spectral function produced by the  errors. However, the low sensitivity of the same quantity to the variation of the strength parameter $\mu$ leads to an overall large spread of the statistical distribution, which explains the larger error quoted in the last column of the first row of Table \ref{tab:analytic}.

For $s_0=m_\tau^2$, the large errors of the input data in the last bins lead to the large uncertainties on $\mu$ quoted in the second line of 
Tab.~\ref{tab:analytic}. In this case, the detection of DVs in a significant way from the pseudodata is not possible.  For the weight (\ref{eq:w}) with  $a=0.96$, the great sensitivity with respect to the input data near the upper end of the spectrum  exceeds the opposite large sensitivity to the variation of $\mu$. The resulting $\mu$ has  a larger uncertainty than that obtained with the standard $L^2$ norm. In the case of  the pinched weight $w_{\rm pd}$,  the suppressing effect on the large errors of the last bins compensates the spread produced by the low sensitivity to $\mu$ variation. The overall effect is that for $s_0=m_\tau^2$ the spreads on $\mu$ obtained with the two extreme weights are comparable. 

A last remark concerns the relation between the weighted $L^2_w$ norms and the $L^\infty$ norm.  One can see that for  $s_0=2.76\gev^2$ the $\mu$ distribution quoted in  Table \ref{tab:analytic} for the weight of the form (\ref{eq:w}) with  $a=0.96$ coincides with that obtained from the distance $\delta_0$ measured in $L^\infty$ norm, given in Table \ref{tab:s0Dep}. However,   for  $s_0=m_\tau^2$, the
standard deviation on $\mu$   quoted in the third column of Table \ref{tab:analytic}  is somewhat smaller than that obtained with $\delta_0$ in Table \ref{tab:s0Dep}. 

To understand this small difference, we recall that the simulations reported in Table \ref{tab:analytic}  were performed with a fixed value,  $a=0.96$, in 
the expression (\ref{eq:w}). But, as discussed in Sec. \ref{sec:approx}, the optimal choice of $a$ achieving the supremum in (\ref{eq:LinfL2}) 
depends also on the input spectral function. For $s_0=2.76\gev^2$, when the bins with large errors are excluded, this dependence affects in an almost unobservable way the simulations. However, for $s_0=m_\tau^2$ the variation of the input can be considerable due to the large errors in the last bins. In this case the weight  (\ref{eq:w}) with  $a=0.96$ is not always the optimal weight leading to the precise approximation of $\delta_0$ according to  Eq.~(\ref{eq:LinfL2}). Therefore, the result presented in Table~\ref{tab:analytic} only illustrates the use of the norm  $L^2_w$ for a rather extreme weight,  inspired from the $L^\infty$ norm but not reproducing exactly its results.  When the errors are large,   the simulations using the $L^\infty$ norm  must resort to the exact algorithm (\ref{eq:del0H}) with the Hankel matrix (\ref{eq:hank}).


\section{Summary and conclusions}
\label{sec:conc}
In the present paper we continued the investigation of the functional-analyses tools proposed in Ref.~\cite{CGP14} for 
detecting DVs from measurements of the spectral functions of the QCD correlators. The aim was to evaluate the potential 
of the method when the spectral function comes in the form of binned data with realistic covariances. We performed the 
analysis still in the context of the toy model for the correlator $\Pi(s)$ considered in \cite{CGP14}, in which we allowed for 
uncertainties described by the covariances obtained from the publicly available ALEPH spectral functions. In this way we 
had full control over the problem and the outcome of the method could be checked against the expected results.

 The paper contains also some theoretical developments of the approach proposed in  Ref.~\cite{CGP14}. In addition to the functional distances based on  $L^2$ and $L^\infty$ norms, already discussed in Ref.~\cite{CGP14},  we introduced a general class of weighted norms $L^2_w$, which are instrumental for several reasons. First,  as shown in Sec. \ref{sec:analyt}, we were able to obtain a closed analytic expression for the minimal functional distance $\delta_{2,w}$ measured in this norm,  thereby avoiding truncated sums of Fourier coefficients. Second,  these norms provide an  interpolation between two extreme cases: the pinched weights familiar from phenomenological works, and the opposite  class of weights which, as discussed in Sec. \ref{sec:approx}, provide a good approximation of the functional distance measured in $L^\infty$ norm. 

 To investigate the potential of the method for the detection of DVs we introduced, in the spirit of Ref.~\cite{CGP14}, a strength parameter $\mu$ that quantifies the DV contribution to $\Pi(s)$ according to (\ref{eq:Pimu}), the true value of this parameter being $\mu=1$. As in \cite{CGP14}, we define the optimal $\mu$ as the value that achieves the minimum of the lower bounds on the functional distances $\delta_0$, $\delta_2$ or $\delta_{2,w}$, measured in the norms $L^2$, $L^\infty$ or $L^2_w$, respectively, between the true correlator and its approximant along the circle $|s|=s_0$ in the complex energy plane.

For want of a theoretical statistical interpretation of the minimal distances defined by functional analysis,  we performed an empirical study where fake data on the spectral function have been generated in a number of bins. To mimic the experimental situation, we adopted a multivariate Gaussian distribution with covariances inferred from the ALEPH covariance matrix for the vector channel~\cite{ALEPH}.  
 By simulations with 5,000 different data sets,  we obtained the statistical
 distributions of the optimal  parameter $\mu$, which were, 
 to a very  good approximation, Gaussian.   

 The main results of these investigations are displayed in Tabs.~\ref{tab:NDep},~\ref{tab:s0Dep} and~\ref{tab:analytic}, where 
 we quote the central values given by the  medians and the uncertainties defined by $68$\% confidence levels from the corresponding distributions. One can see that the results based on the truncated computation of the norms converge relatively fast, which make their practical use feasible.
 This could be confirmed using the analytical determination of $\mu$ given in Eq.~(\ref{eq:muopt}), that avoids the necessity of truncating the sums. We investigated in this framework three types of weights: $w=1$, which corresponds to the standard  $L^2$ norm, the expression (\ref{eq:w}) with $a=0.96$, expected to approximate well the $L^\infty$ norm, and the pinched weight (\ref{eq:wpd}).

 We note that in all cases the true value $\mu=1$ of the strength parameter is obtained with high accuracy. Since the theoretical input we use is quite precise, this result represents a good test of the numerical algorithms adopted.  In particular, as discussed in Sec.~\ref{sec:results}, the refined integration rule (\ref{eq:BetterIntegral}) for calculating either the moments (\ref{eq:cn}) or the quantity (\ref{eq:FDV}) must be used for reaching this level of accuracy.  On the other hand, the standard deviations, crucial for the extraction of DVs in a significant way, differ for various tests.  For the lower value of $s_0$, the tests based on the norms $L^2$ and $L^\infty$ produce comparable uncertainties, with a successful and statistically significant (by three standard deviations) detection of DVs. The test based on pinched weight (\ref{eq:wpd}) is however unable to detect DVs in a significant way even at low $s_0$.  For $s_0=m_\tau^2$, due to the large uncertainties towards the edge of the spectrum, a statistically significant determination of $\mu$ is not possible. All the tests have very large uncertainties, although one may note that the performance of the $L^2$ norm is superior to those of the $L^\infty$ norm and the weighted $L^2_w$ norms with the other two weights considered in Table \ref{tab:analytic}.

 One might ask what is the relation of the present approach to the standard $\chi^2$-type fits used up to now for the phenomenological determination of DVs.  The coefficients $c_n$ defined in Eq.~(\ref{eq:cn}) are actually the moments used in traditional finite-energy sum rules based on a Cauchy integral relation for the correlator $\Pi(s)$ multiplied with a power of $s$ along the contour of Fig. \ref{fig:circle}.  Replacing the approximant $\Pi_{\rm QCD}$ by the exact $\Pi$ and using the exact spectral function $\sigma$, we would have $c_n=0$, by analyticity.  In practice, the coefficients $c_n$ are {\it not} zero due to the imperfections of $\Pi_{\rm QCD}(s)$ and to the statistical fluctuations of experimental values of $\sigma(s)$.

In the standard analyses, starting from this remark a few moments  are selected and combined for defining a certain ``fit quality",  usually a $\chi^2$, with an assumed statistic distribution. This allows, by standard techniques of $\chi^2$ minimization, 
 the extraction of the parameters of the DV models and their covariances  together with the values of other parameters of the OPE, in particular 
 the strong coupling constant and  condensates, for example. The limitation of this approach is that only a small
 number of low-order moments, with known errors and possible correlations, can be included in the fit, 
 due to the fact that in QCD only a small number of power corrections are available. The inclusion of high-order moments must be avoided, as it would introduce unknown high-order condensates~\cite{BGMP16}.
 
 In the present approach, the fact that only a finite number of power corrections are known in the OPE is not essential, because no assumption about the vanishing of specific moments is made. The method exploits the obvious remark that the exact correlator and its approximate representation provided by the QCD calculations available at present are different functions, with different analytic properties. Therefore, the functional distance between them, measured in a certain norm, must exceed a rigorous lower bound. As seen from the algorithms of calculating this lower bound,  all the moments contribute to it, irrespective of the number of power corrections available in the OPE.  The individual moments are actually not relevant, since the minimal functional distance measured in the general $L^2_w$ norm is proved to have the analytic expression (\ref{eq:delta2w}) directly in terms of the spectral function, which obviates the need for the computation of the moments that appear in Eq.~(\ref{eq:cn}).  

Of course, there is a price
to pay, and this is the fact, already
mentioned above, that the functional distances lack a definite statistical interpretation. Therefore, 
in order to extract optimal parameters with definite confidence intervals, one must resort to simulations based on 
fake-data generation. Related to that point is the fact that  the data covariances
enter the procedure solely in the Monte Carlo error propagation and do not affect the evaluation of the functional distances (as they would in a $\chi^2$ analysis).
The extraction of several free parameters in this framework is feasible in principle, but may be complicated in practice. Therefore, in this work we applied the functional approach for the extraction of only one parameter, the strength $\mu$ of the DV term, assuming that all the other parameters are known. The results show that, restricting $s_0$ to values slightly below $m_\tau^2$, in order to avoid the large errors near the end-point of the spectrum in $\tau$ decays, the functional approach is able to detect, in a statistically significant way, the presence of DVs in realistic spectral function pseudodata.

The present analysis paves the road for the next step, of testing DV models with real data. The first task will be to detect
 DVs in the real data employing the formalism described here. This can be carried out along the lines of our study with pseudodata,   with the help of a strength parameter and using values of additional  QCD parameters extracted from other processes. The method can also be
 used to compare different models and, although technically more challenging, also allows for the extraction of parameters entering the models. Having in view the importance of detecting and describing DVs in a reliable way, this framework, although somewhat limited, is of interest as a complementary approach to other phenomenological studies.

\section*{Acknowledgements}

The authors would like to express special thanks to the Mainz Institute for Theoretical Physics (MITP) for its hospitality and support during the commencement of this work. We would like to thank Maarten Golterman and Santi Peris for the careful reading of the manuscript. DB thanks the kind hospitality of IFAE and the Universitat Aut\`onoma de Barcelona where this work was finalized. DB's work was supported by the S\~ao Paulo Research Foundation (FAPESP) grant 2015/20689-9, by CNPq grant 305431/2015-3, and by the Alexander von Humboldt Foundation. IC's work was supported by ANCS, Contract PN 16 42 01 01/2016.

\end{document}